\documentclass[conference]{IEEEtran} 
\IEEEoverridecommandlockouts
\usepackage{cite}
\usepackage{url}
\usepackage{amsmath,amssymb,amsfonts}
\usepackage{algorithm}
\usepackage{algorithmic}
\usepackage{graphicx}
\usepackage{float}
\usepackage{subfig}
\graphicspath{{pics/}}
\newcommand{\figref}[1]{Fig.~\ref{#1}}
\usepackage{makecell}
\usepackage{threeparttable}
\usepackage{textcomp}
\usepackage{xcolor}
\usepackage{fancyhdr}
\def\BibTeX{{\rm B\kern-.05em{\sc i\kern-.025em b}\kern-.08em
    T\kern-.1667em\lower.7ex\hbox{E}\kern-.125emX}}
\begin{document}
\fancyhf{}
\pagestyle{empty}
\thispagestyle{empty}

\title{RIS-Aided Receive Generalized Spatial Modulation Design with Reflecting Modulation\\
\thanks{This paper is supported in part by National Natural Science Foundation of China Program(62371291, 62271316, 62101322), National Key R\&D Project of China (2023YFF0904603), the Fundamental Research Funds for the Central Universities and Shanghai Key Laboratory of Digital Media Processing (STCSM 18DZ2270700). 

The corresponding author is Yin Xu (e-mail: xuyin@sjtu.edu.cn).}
}

\author{

    \IEEEauthorblockN{Xinghao Guo\IEEEauthorrefmark{1}, Yin Xu\IEEEauthorrefmark{1}, Hanjiang Hong\IEEEauthorrefmark{1}, De Mi\IEEEauthorrefmark{2}, Ruiqi Liu\IEEEauthorrefmark{3}, Dazhi He\IEEEauthorrefmark{1}, \\Wenjun Zhang\IEEEauthorrefmark{1}, \textit{Fellow, IEEE}, and Yi-yan Wu\IEEEauthorrefmark{4}, \textit{Life Fellow, IEEE}}
    \IEEEauthorblockA{
    \IEEEauthorrefmark{1} Cooperative Medianet Innovation Center (CMIC), Shanghai Jiao Tong University, Shanghai 200240, China \\
    \IEEEauthorrefmark{2} School of Computing and Digital Technology, Birmingham City University, UK \\  \IEEEauthorrefmark{3} Wireless and Computing Research Institute, ZTE Corporation, Beijing 100029, China \\ \IEEEauthorrefmark{4} Wireless Technology Research Department, Communications Research Centre, Ottawa ON K2K 2Y6, Canada \\ Email: \{guoxinghao@sjtu.edu.cn, xuyin@sjtu.edu.cn, honghj@sjtu.edu.cn, de.mi@bcu.ac.uk, richie.leo@zte.com.cn, \\
    hedazhi@sjtu.edu.cn,  zhangwenjun@sjtu.edu.cn, yiyan.wu@ieee.org\}
    }
    
}

\maketitle

\begin{abstract}
     Spatial modulation (SM) transmits additional information bits by the selection of antennas. Generalized spatial modulation (GSM), as an advanced type of SM, can be divided into diversity and multiplexing (MUX) schemes according to the symbols carried on the selected antennas are identical or different. Recently, reconfigurable intelligent surface (RIS) assisted SM exhibits better reception performance compared to conventional SM. To overcome the limitations of SM, this paper combines GSM with RIS and proposes the RIS-aided receive generalized spatial modulation (RIS-RGSM) scheme. The RIS-RGSM diversity scheme is realized via a simple improvement based on the state-of-the-art scheme. To further increase the transmission rate, a novel RIS-RGSM MUX scheme is proposed, where the reflection phase shifts and on/off states of RIS elements are configured to achieve bit mapping. The theoretical bit error rate (BER) of the proposed scheme is derived and agrees well with the simulation results. Numerical simulations show that the RIS-RGSM MUX scheme has better BER performance than the diversity scheme. The proposed scheme can significantly increase the transmission rate and maintain good performance compared to the existing scheme under a limited number of antennas.
\end{abstract}

\begin{IEEEkeywords}
    RIS, SM, GSM, spectral efficiency, multiplexing.
\end{IEEEkeywords}

\section{Introduction} \label{i}
\thispagestyle{empty}
\IEEEPARstart{W}{ith} the massive growth of mobile devices, future wireless communication is anticipated to attain higher spectral efficiency (SE). Various solutions emerge in the physical layer for enhanced efficiency and reliability, e.g., non-orthogonal multiple access (NOMA) \cite{NOMA-0}, massive multiple-input-multiple-output (MIMO) \cite{MIMO-0}, index modulation (IM) \cite{Qingqing}, reconfigurable intelligent surface (RIS) \cite{Ruiqi}, etc.

As a prominent example of IM techniques, spatial modulation (SM) conveys additional spatial information bits via the index of activated selected transmit antenna \cite{SM}. 
Space shift keying (SSK) is a special form of spatial modulation (SM) in which information bits are transmitted only by selecting the transmit antenna without digital modulation \cite{SSK}. Subsequently, the fundamental SM and SSK are extended to many variants. Receive spatial modulation (RSM) transmits the spatial bits via the index of the receive antenna, which maximizes the signal-to-noise ratio (SNR) at the selected receive antenna with transmitter preprocessing \cite{RSM}. However, the SM, SSK, and RSM schemes all have the following limitations: the number of antennas needs to be a power of 2, and the transmission rate grows logarithmically rather than linearly with the number of antennas. Generalized spatial modulation (GSM), generalized space shift keying (GSSK), and receive generalized spatial modulation (RGSM) have been proposed to alleviate the limitations above  \cite{GSM,GSSK}. GSM/RGSM can be classified into diversity or multiplexing (MUX) schemes based on whether the modulated symbols carried on the selected transmit/receive antennas are identical or different. 

RIS consists of numerous passive elements that manipulate channel scattering and propagation characteristics by introducing pre-designed phases to incident waves \cite{RIS-ruiqi}. RIS can function as the modulator to realize a low-cost and energy-efficient RF chain-free transmitter \cite{RIS-MOD}. RIS can also achieve passive information transmission by controlling the on/off states of its elements \cite{RIS-ONOFF}. Specifically, \cite{RIS-RSM} innovatively integrates RIS with SM, proposing the RIS-aided receive space shift keying (RIS-RSSK) and RIS-aided RSM (RIS-RSM) schemes. In \cite{RIS-RSM}, RIS operates in RIS-access point (RIS-AP) mode, serving as a transmitter module to reflect signals to the selected receive antenna based on spatial bits.
Numerical results demonstrate the superior bit error rate (BER) performance of the RIS-aided schemes compared to the conventional schemes.
\cite{RIS-RGSSK} combines GSSK with RIS and proposes the RIS-aided receive generalized space shift keying (RIS-RGSSK) scheme to mitigate the limitations of SSK. To attain a higher transmission rate, GSM should be introduced into RIS-aided scenarios where the number of receive antennas is limited.

To fill this gap, this paper integrates GSM with RIS and proposes the RIS-aided receive generalized spatial modulation (RIS-RGSM) schemes, encompassing diversity and MUX scheme. The RIS-RGSM diversity scheme can be easily obtained based on the scheme in \cite{RIS-RGSSK}. 
Inspired by \cite{RIS-MOD,RIS-ONOFF}, we propose a novel RIS-RGSM MUX scheme, where the reflection phase shifts and on/off states of RIS elements are manipulated to achieve phase and amplitude modulation of information bits, respectively. Moreover, the theoretical BER of the proposed scheme is analyzed and derived.

\section{RIS-Aided Receive Generalized Spatial Modulation} \label{ii}
\subsection{System Model}
\begin{figure}[htbp]
	\centering
	\subfloat[Diversity scheme\label{diversity}]{
	\includegraphics[width=0.95\columnwidth]{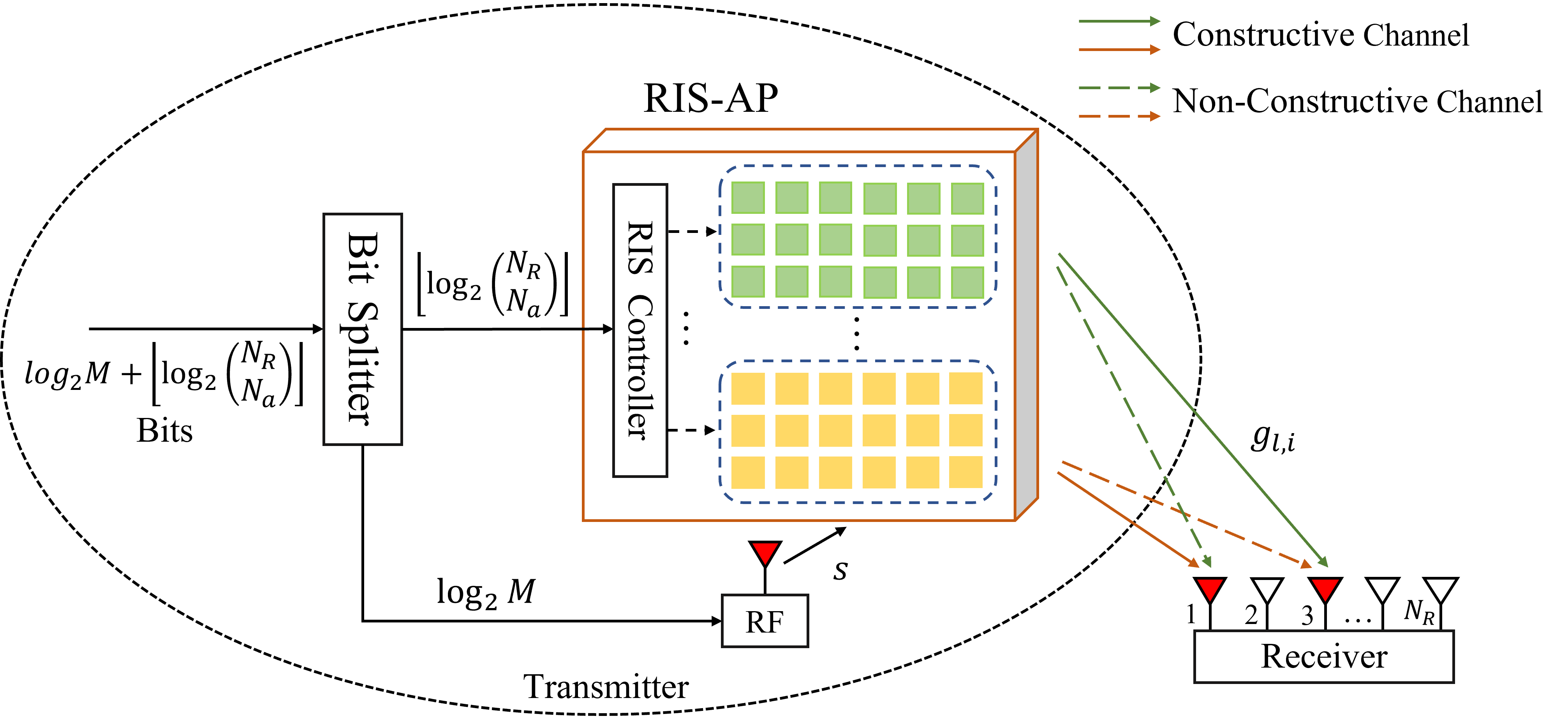}
	}\\
	\subfloat[MUX scheme\label{MUX}]{
	\includegraphics[width=0.99\columnwidth]{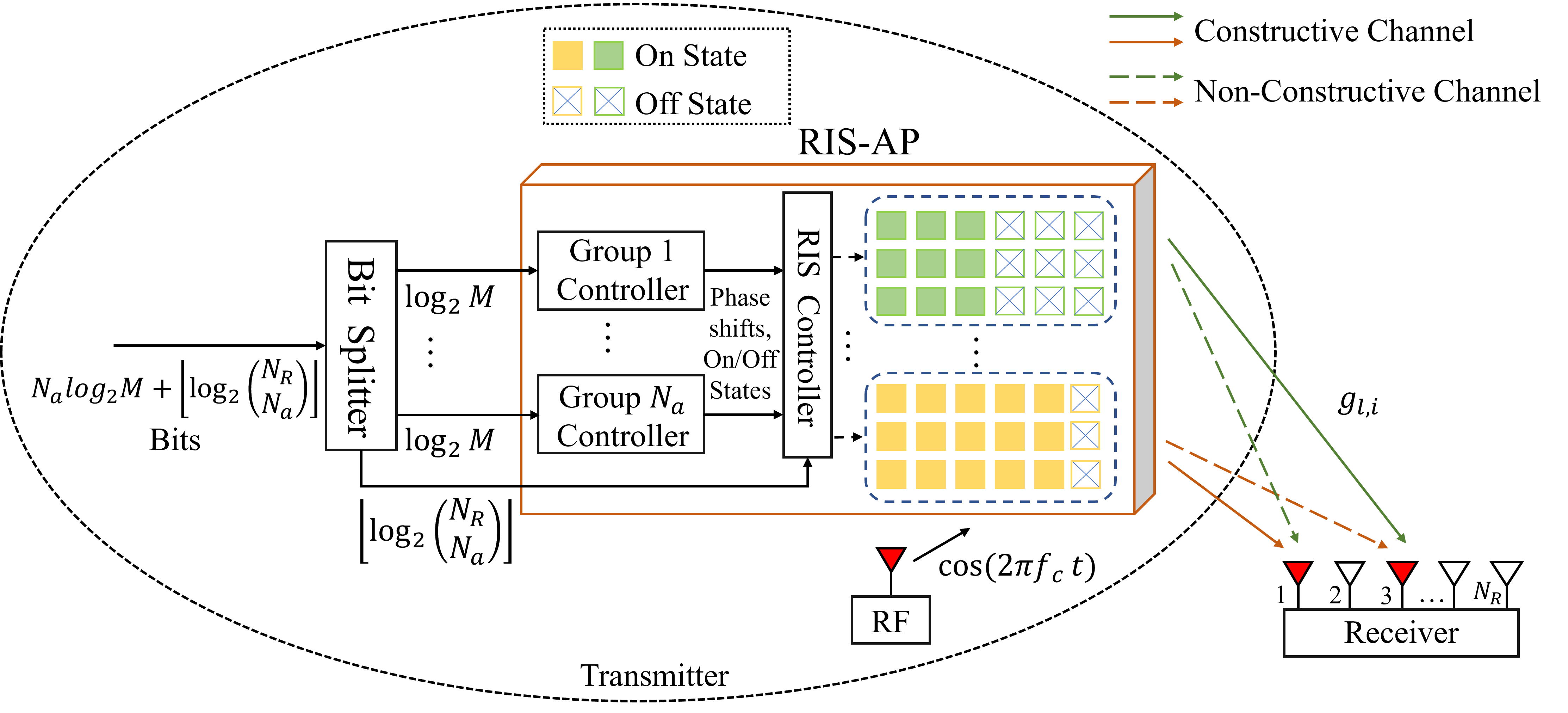}
	}
	\caption{Diagram of the RIS-RGSM system.}
	\label{system}
\end{figure}

The proposed RIS-RGSM systems are shown in \figref{system}, where the transmitter has one antenna, the receiver has $N_{R}$ antennas. 
The RIS with $N$ elements works in the RIS-AP mode,
i.e., as a transmitter module to reflect signals to the receiver. Consistent with \cite{RIS-RSM}, path loss and scattering of between RIS and transmit antenna are negligible \cite{RIS-AP2}, and the RIS controller is assumed to possess the channel phase knowledge between RIS and receive antennas. 

At the receiver, spatial bits are mapped to the index of  the receive antenna combination, which has $N_{a}$ selected antennas with the condition of $N_{a}\leq \lfloor N_{R}/2\rfloor$. The total number of combinations that select $N_{a}$ antennas from $N_{R}$ antennas is $\binom{N_{R}}{N_{a}}$. According to the operational principle of GSM, the number of combinations available for mapping must be a power of two \cite{GSM}. Therefore, it is required to choose $N_{c}=2^{m_0}$ combinations from all combinations, where $m_0$ represents the number of bits corresponding to a combination and given by $m_0=\lfloor\log_2\binom{N_{R}}{N_{a}}\rfloor$. Let $c$ denote the set containing indices of selected antennas corresponding to a combination, i.e., $c=\{c_1,c_2,...,c_{N_{a}}\}$ and $c_l\in\{1,2,...,N_{R}\}$ for $l=1,2,...,N_{a}$. For example, Table \ref{tb:GSM} provides the case when $N_{R}=4$ and $N_a=2$. At each time slot, the receiver decodes spatial bits by determining the combination.
\begin{table}[htbp]
\renewcommand\arraystretch{1.0}
    \centering
    \caption{Antennas index mapper for $N_{R}=4$ and $N_a=2$.}
    \begin{tabular}{|c|c|}
    \hline
        \makecell{$m_0$ bits} & \makecell{$c$} \\ \hline
        00  & \{1,3\}  \\ \hline
        01  & \{1,4\}  \\ \hline
        10  & \{2,3\}  \\ \hline
        11  & \{2,4\} \\ \hline
    \end{tabular}
    \label{tb:GSM}
\end{table}

The RIS controller dynamically adjusts the reflection phases of elements based on input spatial bits and known channel phase knowledge. RIS elements are divided into $N_{a}$ groups, each comprising $N_g=\frac{N}{N_{a}}$ elements. The reflection phase of the $k$-th element in the $l$-th group is denoted as $\phi_{l,k}$, $k\in\{1,2,...,N_g\}$. The reflection phases $\phi_{l,:}$ of the $l$-th group are adjusted to maximize the SNR of the $l$-th antenna in $c$. The reflection coefficients of the elements in the $l$-th group are stored in the column vector $\bold{p}_{l}=[e^{j\phi_{l,1}},e^{j\phi_{l,2}},...,e^{j\phi_{l,N_g}}]^T$.

The wireless channel between the RIS element and the receive antenna is characterized by $g = \beta e^{-j \psi}$ and follows a $\mathcal{CN}(0,1)$ distribution under the assumption of flat Rayleigh fading channels. The channel matrix is denoted as $\bold{G} \in \mathbb{C}^{N_R \times N}$. The $n$-th row of matrix $\bold{G}$ can be segmented into $N$ row vectors, where the $l$-th vector consists of the channel coefficients from elements in the $l$-th group to the $n$-th receive antenna, denoted by $\bold{g}_{nl}= [g^{nl}_1,g^{nl}_2,...,g^{nl}_{N_g}]$. Therefore, the $n$-th row of $\bold{G}$ satisfies $\bold{G}(n,:)= [\bold{g}_{n1}, \bold{g}_{n2},...,\bold{g}_{nN_g}]$ and the $k$-th element of $\bold{g}_{nl}$ satisfies $g^{nl}_k= \bold{G}(n,(l-1)N_g+k) =\beta_{n,(l-1)N_g+k} e^{-j \psi_{n,(l-1)N_g+k}}$. For clarity, the function $f(x,y)$ is defined as $f(x,y)=(x-1)N_g+y$, which gives
\begin{small}
\begin{equation}
\label{channel}
g^{nl}_k= \bold{G}(n,f(l,k))=\beta_{n,f(l,k)} e^{-j \psi_{n,f(l,k)}}.
\end{equation}
\end{small}According to \cite{RIS-RSM}, the $N_R$-dimensional received signal vector $\bold{y}$ can be expressed as
\begin{equation}
\label{y_receive}
\begin{aligned}
\bold{y}&=\sqrt{E_{s}}\bold{G}\bold{p}+\bold{w},\\
\end{aligned}
\end{equation}
where $E_{s}$ is the transmitted signal energy of the unmodulated carrier, $\bold{p}=[\bold{p}_{1}^T,\bold{p}_{2}^T,...,\bold{p}_{N_a}^T]^T$, and $\bold{w}$ is the additive white Gaussian noise (AWGN) with zero mean and variance $\sigma^2 \bold{I}_{N_R}$.

\subsection{Transmission Schemes}
The transmission of RIS-RGSM is divided into diversity scheme and MUX scheme. In the context of MUX scheme, the phase shift keying (PSK) is used as an example of how to implement phase modulation, while the amplitude phase shift keying (APSK) is used to illustrate the simultaneous modulation of amplitude and phase.
\subsubsection{RIS-RGSM diversity scheme}
In \cite{RIS-RGSSK}, if the $n$-th antenna of the receiver is selected as the $l$-th antenna in the combination $c$, i.e. $c_l=n$, the reflection phase of the $k$-th element in the $l$-th group should satisfy \begin{small}
\begin{equation}
\label{phase_gsm_div}
\phi_{l,k}=\psi_{n,f(l,k)},
\end{equation}
\end{small}assuming the RIS has the knowledge of channel phases. 

Incorporating the process of the transmitter sending modulated symbols on the foundation of \cite{RIS-RGSSK}, the RIS-RGSM diversity scheme can be realized, as shown in \figref{diversity}. The received signal at the $n$-th antenna can be expressed as:
\begin{small}
\begin{equation}
\label{gsm_div_receive}
y_n=\sqrt{E_{s}}\Bigg(\sum_{k=1}^{N_g}\beta_{n,f(l,k)}+\!\!\!\!\sum_{i=1,i\neq l}^{N_a}\bold{g}_{ni}\bold{p}_{i}\Bigg)s+w_n,
\end{equation}
\end{small}where $\bold{g}_{ni}\bold{p}_{i}=\sum_{k=1}^{N_g}\bold{G}(n,f(i,k))e^{j\psi_{c_i,f(i,k)}}$, and $s$ denotes the modulated symbol. According to \eqref{gsm_div_receive}, it is observed that the received signal of the $l$-th selected antenna is divided into two components. The first component is constructive after the reflection phases of the $N_g$ elements in the $l$-th group are configured to cancel out the channel phases. The second component, comprising signals reflected from elements belonging to other groups, is non-constructive and consequently treated as interference. Hence, the signal quality of the selected antenna is enhanced due to the phase alignment of the reflected signals from the corresponding group elements. 

In the RIS-RGSM diversity scheme, a portion of the information bits is transmitted by selecting different combinations of receive antennas, whereas the remaining portion is carried by modulated symbols. Therefore, the transmission rate is $R=m_0+\log_2M$ bits per channel use (bpcu), with the modulation order $M$. Meanwhile, the rate of the RIS-RGSSK scheme, as discussed in \cite{RIS-RGSSK}, is only $m_0$ bpcu.

\subsubsection{RIS-RGSM MUX scheme with PSK}
Compared with the RIS-RGSM diversity scheme, in the MUX scheme, the transmit antenna only serves as an excitation, while the information bits originally mapped to modulated symbols are alternatively transmitted via the control of RIS elements. 


As shown in \figref{MUX}, each RIS group is correspondingly equipped with a controller to independently acquire an additional phase $\varphi_l$ based on $m_1=\log_2M$ information bits, where $M$ is the equivalent order of the received constellation, thus $\varphi_l\in\{0,\frac{2\pi}{M},...,\frac{2\pi(M-1)}{M}\}$. 
So if the $n$-th antenna of the receiver is selected as the $l$-th antenna in the combination $c$, the reflection phase of the $k$-th element in the $l$-th group is given by
\begin{equation}
\label{phase_gsm}
\phi_{l,k}=\psi_{n,f(l,k)}+\varphi_l.
\end{equation}
The received signal can be re-formulated as
\begin{small}
\begin{equation}
\label{gsm_psk_receive}
\begin{aligned}
y_n=\sqrt{E_{s}}\Bigg(\sum_{k=1}^{N_g}\beta_{n,f(l,k)}e^{j\varphi_l}+\!\!\!\!\sum_{i=1,i\neq l}^{N_a}\bold{g}_{ni}\bold{p}_{i}\Bigg)+w_n.
\end{aligned}
\end{equation}
\end{small}Compared with \eqref{gsm_div_receive}, the $l$-th selected antenna equivalently receives the exclusive symbol $s_l=\sqrt{E_{s}}e^{j\varphi_l}$, and the transmission rate increases to
\begin{equation}
\label{bpcu}
R=m_0 + N_a m_1 \quad \text{bpcu}.
\end{equation}

\subsubsection{RIS-RGSM MUX scheme with APSK}
To change the amplitude of the equivalent received symbol $s_l$ at the $l$-th selected antenna, the RIS group controller can modify the number of activated (ON state) elements within the group, as shown in \figref{MUX}. 

For APSK with modulation order $M=M_rM_p$, constellation symbols are scattered on $M_r$ concentric rings with uniformly increasing radius and each ring contains $M_p$ uniform-distributed constellation symbols. The $m_1$ bits can be divided into two parts, $m_p=\log_2M_p$ and $m_r=\log_2M_r$. The $m_p$ bits are mapped to the additional phase $\varphi_l$, and $m_r$ bits are mapped to the number of activated elements $a_l$, where $\varphi_l \in \{0,\frac{2\pi}{M_p},...,\frac{2\pi(M_p-1)}{M_p}\}$ and $a_l \in \{\frac{N_g}{M_r},\frac{2N_g}{M_r},...,N_g\}$. Then the reflection coefficient vector $\bold{p}_{l}$ of the $l$-th group satisfies $\bold{p}_{l}=[e^{j\phi_{l,1}},e^{j\phi_{l,2}},...,e^{j\phi_{l,a_l}},0,...,0]^T$, $a_l \in \{\frac{N_g}{M_r},\frac{2N_g}{M_r},...,N_g\}$. The received signal is written as
\begin{small}
\begin{equation}
\label{gsm_apsk_receive}
\begin{aligned}
y_n=\sqrt{E_{s}}\Bigg(\sum_{k=1}^{a_l}\beta_{n,f(l,k)}e^{j\varphi_l}+\!\!\!\!\sum_{i=1,i\neq l}^{N_a}\bold{g}_{ni}\bold{p}_{i}\Bigg)+w_n.
\end{aligned}
\end{equation}
\end{small}When $\frac{N_g}{M_r}\gg1$, applying the central limit theorem (CLT), the equivalent received symbol $s_l$ can be approximated as
\begin{equation}
\label{symbol}
s_l=\sqrt{E_{s}}\rho_l e^{j\varphi_l},
\end{equation}
where $\rho_l\in\{\frac{1}{M_r},\frac{2}{M_r},...,1\}$. So the received signal can be approximated as
\begin{small}
\begin{equation}
\label{gsm_apsk_receive_app}
\begin{aligned}
y_n=&\sum_{k=1}^{N_g}\beta_{n,f(l,k)}s_l+\!\sqrt{E_{s}}\!\!\!\sum_{i=1,i\neq l}^{N_a}\bold{g}_{ni}\bold{p}_{i}+w_n\\
=&\sum_{k=1}^{N_g}\beta_{n,f(l,k)}s_l+\!\!\sum_{i=1,i\neq l}^{N_a}\sum_{k=1}^{N_g}\bold{G}(n,f(i,k))e^{j\psi_{c_i,f(i,k)}}s_i+w_n.
\end{aligned}
\end{equation}
\end{small}

\addtolength{\topmargin}{0.04in}
\subsection{Detection Criterion}
Assuming that the receiver has complete channel knowledge, the principle of Maximum Likelihood (ML) detection can be expressed as
\begin{small}
\begin{equation}
\begin{aligned}
    \label{ML}
    (\hat{c},\hat{\mathbf{s}})&=\underset{c,\mathbf{s}}{\arg \min }\sum_{n=1}^{N_R}\!\Bigg|y_n-\!\! \sum_{i=1}^{N_a}\sum_{k=1}^{N_g}\bold{G}(n,f(i,k))e^{j\psi_{c_i,f(i,k)}}s_i\Bigg|^2\\
    &=\underset{c,\mathbf{s}}{\arg \min }\sum_{n=1}^{N_R}\!\Bigg|y_n- \sum_{i=1}^{N_a}h_{ni}s_i\Bigg|^2\\
    &=\underset{c,\mathbf{s}}{\arg \min }\sum_{n=1}^{N_R}\!\Big|y_n-\mathbf{h}_{n}\cdot\mathbf{s}\Big|^2,
\end{aligned}
\end{equation}
\end{small}where $h_{ni}=\sum_{k=1}^{N_g}\bold{G}(n,f(i,k))e^{j\psi_{c_i,f(i,k)}}$, the row vector $\mathbf{h}_n=[h_{n1},h_{n2},...,h_{nN_a}]$ and the column vector $\mathbf{s}=[s_1,s_2,...,s_{N_a}]^T$, which comprises $N_a$ equivalent received symbols. 

\section{Performance Analysis}  \label{iii}
This section analyzes the theoretical bit error rate (BER) of the RIS-RGSM MUX system. The theoretical BER of the RIS-RGSM diversity system can be referred to in \cite{RIS-RGSSK}, with minor modifications required.

From \eqref{ML}, conditioned on channel coefficients, the pairwise error probability (PEP) can be expressed as:
\begin{small}
\begin{equation}
\begin{aligned}
    \label{PEP}
    &P(c,\mathbf{s}\to\hat{c},\hat{\mathbf{s}} |\mathbf{G})\\
    &=P\Big(\sum_{n=1}^{N_R}\!\big|y_n-\mathbf{h}_{n}\cdot\mathbf{s}\big|^2> \sum_{n=1}^{N_R}\!\big|y_n-\hat{\mathbf{h}}_{n}\cdot\hat{\mathbf{s}}\big|^2\Big)\\
    &=P\Big(-\sum_{n=1}^{N_R}\big|\mathbf{h}_{n}\mathbf{s}-\hat{\mathbf{h}}_{n}\hat{\mathbf{s}}\big|^2- 2\Re\Big\{\sum_{n=1}^{N_R}n_n^*\big(\mathbf{h}_{n}\mathbf{s}-\hat{\mathbf{h}}_{n}\hat{\mathbf{s}}\big)\Big\}>0\Big)\\
    &=P(B>0).
\end{aligned}
\end{equation}
\end{small}Here, $B\sim\mathcal{N}(\mu_B,\sigma^2_B)$ with $\mu_B=-\sum_{n=1}^{N_R}\big|\mathbf{h}_{n}\mathbf{s}-\hat{\mathbf{h}}_{n}\hat{\mathbf{s}}\big|^2$ and $\sigma^2_B=2N_RN_0\big|\mathbf{h}_{n}\mathbf{s}-\hat{\mathbf{h}}_{n}\hat{\mathbf{s}}\big|^2$. From $P(B>0)=Q(\sqrt{\mu_B^2/\sigma_B^2})$, it follows that 
\begin{small}
\begin{equation}
\begin{aligned}
    \label{QFun}
    P(c,\mathbf{s}\to\hat{c},\hat{\mathbf{s}}|\mathbf{G})
    &=Q\left(\sqrt{\frac{\sum_{n=1}^{N_R}\big|\mathbf{h}_{n}\mathbf{s}-\hat{\mathbf{h}}_{n}\hat{\mathbf{s}}\big|^2}{2N_0}}\right)\\
    &= Q\left(\sqrt{\frac{\|\bold{D}\|^2}{2N_0}}\right)= Q\left(\sqrt{\frac{\Gamma}{2N_0}}\right),
\end{aligned}
\end{equation}
\end{small}where $\|\cdot\|$ denotes the L2 norm, $\bold{D}=[D_1,D_2,...,D_{N_R}]^T$ with $n$-th element satisfies $D_n= \mathbf{h}_{n}\mathbf{s}-\hat{\mathbf{h}}_{n}\hat{\mathbf{s}}$. 

Based on the upper bound of the Q-function \cite{Qfun}, the average PEP can be expressed as:
\begin{small}
\begin{equation}
\begin{aligned}
    \label{P_av}
    &\bar{P}(c,\mathbf{s}\to\hat{c},\hat{\mathbf{s}})\\
    &=E_{\Gamma}\Bigg[Q\Bigg(\sqrt{\frac{\Gamma}{2N_0}}\Bigg)\Bigg]\\
    &\leq \frac{1}{6}E_{\Gamma}\Big[e^{-\frac{\Gamma}{N_0}}\Big]+\frac{1}{12}E_{\Gamma}\Big[e^{-\frac{\Gamma}{2N_0}}\Big]+\frac{1}{4}E_{\Gamma}\Big[e^{-\frac{\Gamma}{4N_0}}\Big]\\
    &=\frac{1}{6}M_{\Gamma}\Big(\frac{-1}{N_0}\Big)+\frac{1}{12}M_{\Gamma}\Big(\frac{-1}{2N_0}\Big)+\frac{1}{4}M_{\Gamma}\Big(\frac{-1}{4N_0}\Big),
\end{aligned}
\end{equation}
\end{small}where $M_\Gamma(x)=E_{\Gamma}[e^{x\Gamma}]$ is the moment generating function (MGF) of $\Gamma$. Note that $\Gamma=\|\bold{D}\|^2$, define $D_n=D_n^R+jD_n^I$, where $D_n^R=\Re\{\mathbf{h}_{n}\mathbf{s}- \hat{\mathbf{h}}_{n}\hat{\mathbf{s}}\}$ and $D_n^I=\Im\{\mathbf{h}_{n}\mathbf{s}- \hat{\mathbf{h}}_{n}\hat{\mathbf{s}}\}$. Depending on whether the detection is correct or erroneous, $D_n$ can be classified into the following categories:
\subsubsection{\rm The $n$-th receive antenna does not belong to the combination $c$ and decoded correctly, i.e., $c_i\neq n$ and $\hat{c}_i\neq n$ for $i\in \{1,2,...,N_a\}$}  \label{iii-1}
\begin{small}
\begin{equation}
\begin{aligned}
    \label{Dn1}
    D_n=\sum_{i=1}^{N_a}\sum_{k=1}^{N_g}\bold{G}(n,f(i,k))(e^{j\psi_{c_i,f(i,k)}}s_i\!-\!e^{j\psi_{\hat{c}_i,f(i,k)}}\hat{s_i}).
\end{aligned}
\end{equation}
\end{small}Here, according to $E[\beta e^{-j\psi}]=0$, $Var[\beta e^{-j\psi}]=1$ \cite{RIS-RSM}, and the CLT, $D_n^R$ and $D_n^I$ follow the same Gaussian distribution of $\mathcal{N}\big(0,\frac{N_g}{2}(\|\mathbf{s}\|^2+ \|\hat{\mathbf{s}}\|^2)\big)$.

\subsubsection{\rm The $n$-th receive antenna is selected as the $l$-th antenna in $c$ and decoded correctly, i.e., $c_l=\hat{c}_l=n$}  \label{iii-2}
\begin{small}
\begin{equation}
\begin{aligned}
    \label{Dn2}
    D_n&=\sum_{k=1}^{N_g}\beta_{n,f(l,k)}(s_l-\hat{s_l})+\\
    &\sum_{i=1,i\neq l}^{N_a}\sum_{k=1}^{N_g}\bold{G}(n,f(i,k))(e^{j\psi_{c_i,f(i,k)}}s_i\!-\!e^{j\psi_{\hat{c}_i,f(i,k)}}\hat{s_i})\\
    &=d_1+d_2.
\end{aligned}
\end{equation}
\end{small}According to the category \textit{1)}, $d_2^R$ and $d_2^I$ follow the same Gaussian distribution of $\mathcal{N}\big(0,\frac{N_g}{2}\sum_{i=1,i\neq l}^{N_a}(|s_i|^2+|\hat{s}_i|^2)\big)$. Besides, based on $E[\beta]=\frac{\sqrt{\pi}}{2}$ and $Var[\beta]=\frac{4-\pi}{4}$ \cite{RIS-RSM}, $\mu_{d_1^R}=\frac{N_g\sqrt{\pi}}{2}(s_l-\hat{s}_l)_{\Re}$, $\mu_{d_1^I}=\frac{N_g\sqrt{\pi}}{2}(s_l-\hat{s}_l)_{\Im}$, $\sigma^2_{d_1^R}=\frac{N_g(4-\pi)}{4}(s_{l}-\hat{s}_{l})^2_{\Re}$, and $\sigma^2_{d_1^I}=\frac{N_g(4-\pi)}{4}(s_{l}-\hat{s}_{l})^2_{\Im}$. Hence, $\mu_{D_n^R}=\frac{N_g\sqrt{\pi}}{2}(s_l-\hat{s}_l)_{\Re}$, $\mu_{D_n^I}=\frac{N_g\sqrt{\pi}}{2}(s_l-\hat{s}_l)_{\Im}$,  $\sigma^2_{D_n^R}=\frac{N_g(4-\pi)}{4}(s_{l}-\hat{s}_{l})^2_{\Re}+\frac{N_g}{2}\sum_{i=1,i\neq l}^{N_a}(|s_i|^2+|\hat{s}_i|^2)$, and $\sigma^2_{D_n^I}=\frac{N_g(4-\pi)}{4}(s_{l}-\hat{s}_{l})^2_{\Im}+\frac{N_g}{2}\sum_{i=1,i\neq l}^{N_a}(|s_i|^2+|\hat{s}_i|^2)$. Unfortunately, although $D_n^R$ and $D_n^I$ both obey Gaussian distributions, they are not independent of each other, and the mean vector and covariance matrix of $[D_n^R,D_n^I]^T$ are as follows:
\begin{small}
\begin{equation}
    \label{Dn2-m}
    \bold{m}=[\mu_{D_n^R},\mu_{D_n^I}]^T,
\end{equation}
\begin{equation}
\begin{aligned}
    \label{Dn2-C}
    \bold{C}=
    \begin{bmatrix}
    \sigma^2_{D_n^R} & \sigma_{D_n^R,D_n^I} \\
    \sigma_{D_n^R,D_n^I} & \sigma^2_{D_n^I}
    \end{bmatrix},
\end{aligned}
\end{equation}
\end{small}where $\sigma_{D_n^R,D_n^I}=\frac{N_g(4-\pi)}{4}(s_{l}-\hat{s}_{l})_{\Re}(s_{l}-\hat{s}_{l})_{\Im}$.

\subsubsection{\rm The $n$-th receive antenna is selected as the $l$-th antenna in $c$ but decoded erroneously, the $m$-th receive antenna is decoded as the $l$-th antenna in $c$, i.e., $c_l=n$ but $\hat{c}_l=m$}  \label{iii-3}
\begin{small}
\begin{equation}
\begin{aligned}
    \label{Dn3}
    D_n&=\sum_{k=1}^{N_g}\beta_{n,f(l,k)}(s_l-e^{j(\psi_{m,f(i,k)}-\psi_{n,f(i,k)})}\hat{s_l})+\\
    &\!\!\sum_{i=1,i\neq l}^{N_a}\!\sum_{\,k=1}^{N_g}\bold{G}(n,f(i,k))(e^{j\psi_{c_i,f(i,k)}}s_i\!-\!e^{j\psi_{\hat{c}_i,f(i,k)}}\hat{s_i})\\
    &=d_1+d_2,
\end{aligned}
\end{equation}
\begin{equation}
\begin{aligned}
    \label{Dm3}
    D_m&=\sum_{k=1}^{N_g}\beta_{m,f(l,k)}(e^{j(\psi_{n,f(i,k)}-\psi_{m,f(i,k)})}s_l-\hat{s_l})+\\
    &\!\!\sum_{i=1,i\neq l}^{N_a}\!\sum_{\,k=1}^{N_g}\!\bold{G}(m,f(i,k))(e^{j\psi_{c_i,f(i,k)}}s_i\!-\!e^{j\psi_{\hat{c}_i,f(i,k)}}\hat{s_i})\\
    &=d_3+d_4.
\end{aligned}
\end{equation}
\end{small}According to the category \textit{2)}, $d_2^R$, $d_2^I$, $d_4^R$ and $d_4^I$ all follow the same Gaussian distribution of $\mathcal{N}\big(0, \frac{N_g}{2}\sum_{i=1,i\neq l}^{N_a}(|s_i|^2+|\hat{s}_i|^2)\big)$. Besides, according to \cite{RIS-RSM}, $\mu_{d_1^R}=\frac{N_g\sqrt{\pi}s_{l\Re}}{2}$, $\mu_{d_1^I}=\frac{N_g\sqrt{\pi}s_{l\Im}}{2}$, $\sigma^2_{d_1^R}=\frac{N_g(4-\pi)s^2_{l\Re}}{4}+\frac{N_g|\hat{s}_{l}|^2}{2}$, and $\sigma^2_{d_1^I}=\frac{N_g(4-\pi)s^2_{l\Im}}{4}+\frac{N_g|\hat{s}_{l}|^2}{2}$. Similarly, the mean and variance of the real and imaginary parts of $D_m$ can be obtained. Hence, the mean vector and covariance matrix of $[D_n^R,D_n^I,D_m^R,D_m^I]^T$ are as follows:
\begin{small}
\begin{equation}
    \label{Dnm3-m}
    \bold{m}_{D_n,D_m}=[\frac{N_g\sqrt{\pi}s_{l\Re}}{2}, \frac{N_g\sqrt{\pi}s_{l\Im}}{2},-\frac{N_g\sqrt{\pi}\hat{s}_{l\Re}}{2}, -\frac{N_g\sqrt{\pi}\hat{s}_{l\Im}}{2}]^T,
\end{equation}
\begin{equation}
\begin{aligned}
    \label{Dnm3-C}
    \bold{C}_{D_n,D_m}=
    \begin{bmatrix}
    \sigma^2_{D_n^R} & \sigma_{D_n^R,D_n^I} & \sigma_{D_n^R,D_m^R} & \sigma_{D_n^R,D_m^I}\\
    \sigma_{D_n^R,D_n^I} & \sigma^2_{D_n^I} & \sigma_{D_n^I,D_m^R} & \sigma_{D_n^I,D_m^I}\\
    \sigma_{D_n^R,D_m^R} & \sigma_{D_n^I,D_m^R} & \sigma^2_{D_m^R} & \sigma_{D_m^R,D_m^I}\\
    \sigma_{D_n^R,D_m^I} & \sigma_{D_n^I,D_m^I} & \sigma_{D_m^R,D_m^I} & \sigma^2_{D_m^I}
    \end{bmatrix},
\end{aligned}
\end{equation}
\end{small}where 
\begin{small}
$\sigma^2_{D_n^R}=\frac{N_g(4-\pi)s^2_{l\Re}}{4}+\frac{N_g\sum_{i=1,i\neq l}^{N_a}|s_i|^2}{2}+\frac{N_g\|\hat{s}\|^2}{2}$,

$\sigma^2_{D_n^I}=\frac{N_g(4-\pi)s^2_{l\Im}}{4}+\frac{N_g\sum_{i=1,i\neq l}^{N_a}|s_i|^2}{2}+\frac{N_g\|\hat{s}\|^2}{2}$,

$\sigma^2_{D_m^R}=\frac{N_g(4-\pi)\hat{s}^2_{l\Re}}{4}+\frac{N_g\sum_{i=1,i\neq l}^{N_a}|\hat{s}_i|^2}{2}+\frac{N_g\|s\|^2}{2}$,

$\sigma^2_{D_m^I}=\frac{N_g(4-\pi)\hat{s}^2_{l\Im}}{4}+\frac{N_g\sum_{i=1,i\neq l}^{N_a}|\hat{s}_i|^2}{2}+\frac{N_g\|s\|^2}{2}$,

$\sigma_{D_n^R,D_n^I}=\frac{N_g(4-\pi)s_{l\Re}s_{l\Im}}{4}$, $\sigma_{D_m^R,D_m^I}=\frac{N_g(4-\pi)\hat{s}_{l\Re}\hat{s}_{l\Im}}{4}$,

$\sigma_{D_n^R,D_m^R}=\frac{N_g\pi(-s_{l\Re}\hat{s}_{l\Re}+s_{l\Im}\hat{s}_{l\Im})}{8}$, $\sigma_{D_n^I,D_m^I}=-\sigma_{D_n^R,D_m^R}$,

$\sigma_{D_n^R,D_m^I}=-\frac{N_g\pi(s_{l\Re}\hat{s}_{l\Im}+\hat{s}_{l\Re}s_{l\Im})}{8}$, $\sigma_{D_n^I,D_m^R}=\sigma_{D_n^R,D_m^I}$.
\end{small}



In the end, based on all the categories of $D_n$ above, the mean vector and covariance matrix of $\Gamma$ are substituted into the following formula:
\begin{small}
\begin{equation}
\begin{aligned}
    \label{MGF}
    M_{\Gamma}(x)=[\det(\mathbf{I}-2x\mathbf{C}_{\Gamma})]^{-\frac{1}{2}} e^{-\frac{1}{2}\mathbf{m}_{\Gamma}^T[\mathbf{I}-(\mathbf{I}-2x\mathbf{C}_{\Gamma})^{-1}]\mathbf{C}_{\Gamma}^{-1}\mathbf{m}_{\Gamma}}.
\end{aligned}
\end{equation}
\end{small}Substituting the MGF in \eqref{MGF} into \eqref{P_av} provides the desired average PEP. Finally, the average BER can be derived as:
\begin{small}
\begin{equation}
    \label{P_b}
    P_b\leq\frac{1}{N_cM^{N_a}}\sum_{c}\sum_{\mathbf{s}}\sum_{\hat{c}}\sum_{\hat{\mathbf{s}}}\frac{\bar{P}(c,\mathbf{s}\to\hat{c},\hat{\mathbf{s}})e(c,\mathbf{s}\to\hat{c},\hat{\mathbf{s}})}{m_0 + N_a m},
\end{equation}
\end{small}where $e(c,\mathbf{s}\to\hat{c},\hat{\mathbf{s}})$ represents the number of bits in error for the corresponding pairwise error event.

\section{Simulation Results}  \label{iv}
This section evaluates the proposed RIS-RGSM system performance and makes some comparisons. The large-scale path loss is not considered since it is implicitly taken into account in the received SNR. Assuming the receiver has complete channel knowledge, the RIS controllers have phase information of the Rayleigh channel.  
For the fairness of comparison, $E_s$ is assumed to be 1 in the simulation.
\begin{figure}[htbp]
    \centerline{\includegraphics[width=0.45\textwidth]{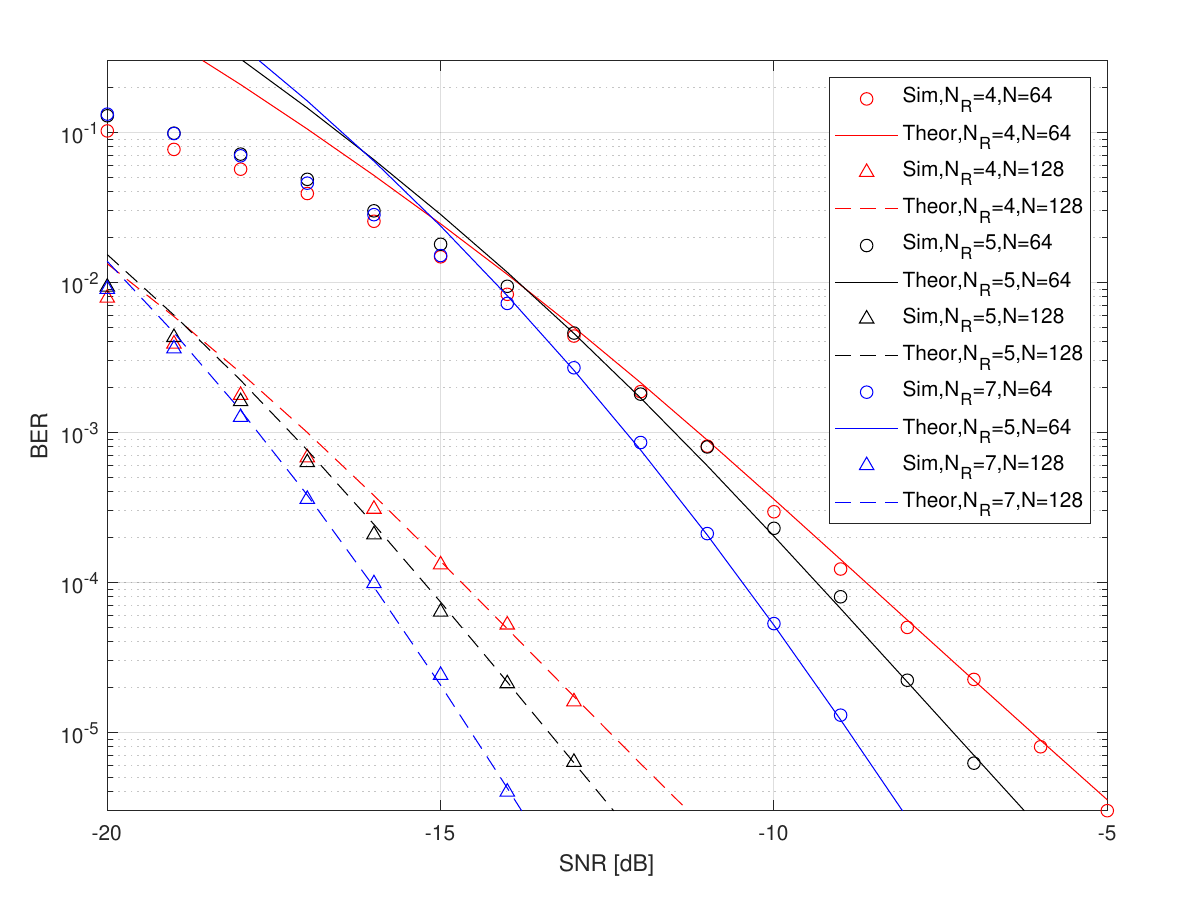}}
    \caption{The BER results of the RIS-RGSM MUX scheme with 8APSK and $N_a=2$ for different conditions in $N$ and $N_R$.}
    \label{fig:BER1}
\end{figure}

First, the impact of changes in antenna configuration on the RIS-RGSM MUX scheme is shown in \figref{fig:BER1}. As expected, the theoretical BER curves derived from the analysis in Section \ref{iii} match well with the results of the Monte Carlo simulation.
Fixing $N_a=2$, vary the number of RIS elements $N$ and the number of receive antennas $N_R$. APSK modulation is adopted with $M_r=2$, $M_p=4$ and $M=M_rM_p=8$. To increase performance, the the reflection phases $\phi_{2,:}$ of the second group all plus a phase offset of $\frac{\pi}{M_p}$, staggered from the equivalent received symbols of the first selected antenna. The same operation can also be applied when $N_a>2$. The results show that when $N_R$ remains constant, the BER performance improves as $N$ increases. When the number of $N$ doubles, the SNR required to achieve the same BER decreases by approximately 6 dB. This is rational because the larger the number of elements allocated to each group, the more reflection paths will be focused on the corresponding selected antenna.
It is worth noting that increasing $N_R$ not only provides a higher transmission rate $R$, but also makes the BER performance better. The phenomenon is consistent with RIS-RSM scheme in \cite{RIS-RSM}. Moreover, although increasing $N$ and $N_R$ can improve the BER performance, the corresponding complexity and hardware overhead will increase. 

\begin{figure}[htbp]
    \centering
	\subfloat[$R=$7\label{fig:BER21}]{
	\includegraphics[width=.315\columnwidth]{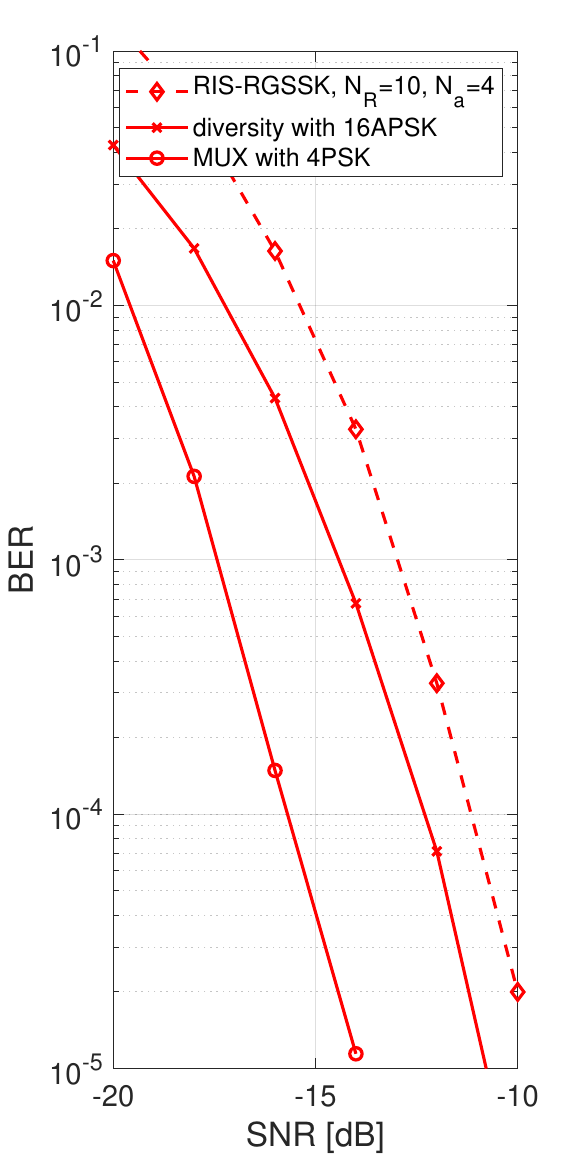}
 	}
	\subfloat[$R=$9\label{fig:BER22}]{
	\includegraphics[width=.315\columnwidth]{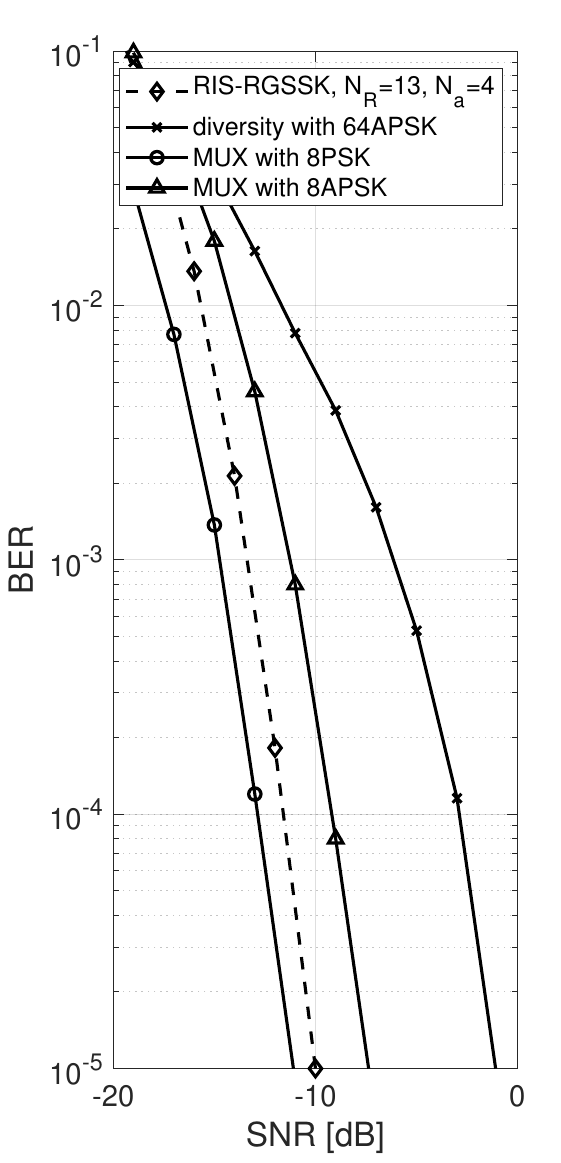}
	}
	\subfloat[$R=$11\label{fig:BER23}]{
	\includegraphics[width=.315\columnwidth]{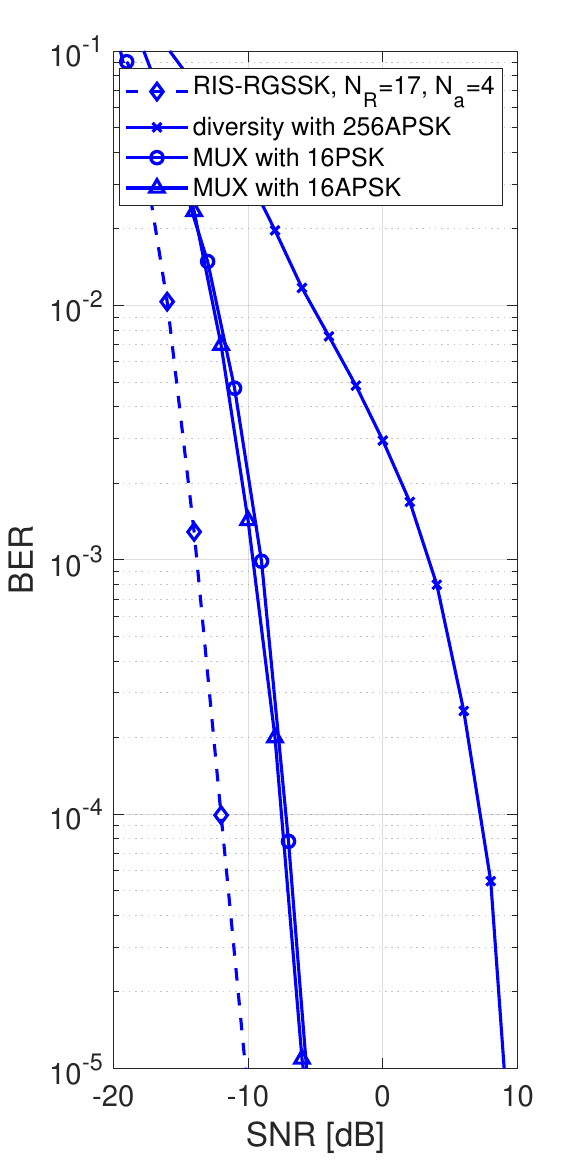}
	}
	\caption{The BER comparison results of RIS-RGSSK with different $N_R,N_a$, RIS-RGSM diversity and MUX with $N_R=5$, $N_a=2$ for $N=64$ and $R=$ (a) 7, (b) 9, (c) 11 bpcu.}
	\label{fig:BER2}
\end{figure}
Next, under the same transmission rate and with $N$=64, the performance comparison results of RIS-RGSM MUX scheme, diversity scheme, and RIS-RGSSK scheme are shown in in \figref{fig:BER2}. For example, in \figref{fig:BER22}, for RIS-RGSM with $N_{R}=5,N_{a}=2$, $m_0=\lfloor\log_2\binom{N_{R}}{N_{a}}\rfloor=\lfloor\log_2\binom{5}{2}\rfloor=3$, the modulation order of the RIS-RGSM diversity scheme is $M_{1}=64$, so the rate is $R_{1}=\log_2{M_{1}}+m_0=6+3=9$ bpcu; the modulation order of RIS-RGSM MUX scheme is $M_{2}=8$, so the rate is $R_{2}=N_a\log_2{M_{2}}+m_0=2\times3+3=9$ bpcu; for RIS-RGSSK with $N_{R}=13,N_{a}=4$, the rate is $R_{3}=m_0 =\lfloor\log_2\binom{10}{4}\rfloor=9$ bpcu. As presented in \figref{fig:BER2}, the BER performance of the RIS-RGSM MUX scheme is better than the diversity scheme when the transmission rate is equal. To achieve the same BER, for the MUX with PSK scheme, the SNR requirement is about 3, 10, and 15 dB lower than the diversity scheme when $R=$7, 9, and 11 bpcu, respectively; for the MUX with APSK scheme, the SNR requirement is about 7 and 15 dB lower than the diversity scheme when $R=$9 and 11 bpcu, respectively. The simulation results show the performance superiority of the proposed RIS-RGSM MUX scheme. The higher the transmission rate, the greater the performance gain of the MUX scheme than the diversity scheme. Besides, the BER performance of the RIS-RGSM MUX scheme is 4.5 and 1.5 dB better than RIS-RGSSK when $R=$7 and 9, respectively. However, as $N_R$ increases, it can be observed that the BER performance of RIS-RGSSK is 4 dB better than the RIS-RGSM MUX scheme for $R=$11. Therefore, in scenarios where the number of receive antennas is limited, RIS-RGSM scheme provides a solution to increase the transmission rate. At the same transmission rate, the RIS-RGSM MUX scheme achieves better performance than the RIS-RGSM diversity scheme at the cost of complexity. Besides, in \figref{fig:BER23}, it can be observed that the performance of the RIS-RGSM MUX with APSK scheme is better than that of MUX with PSK scheme, which leads to the later simulation.

\begin{figure}[htbp]
    \centerline{\includegraphics[width=0.45\textwidth]{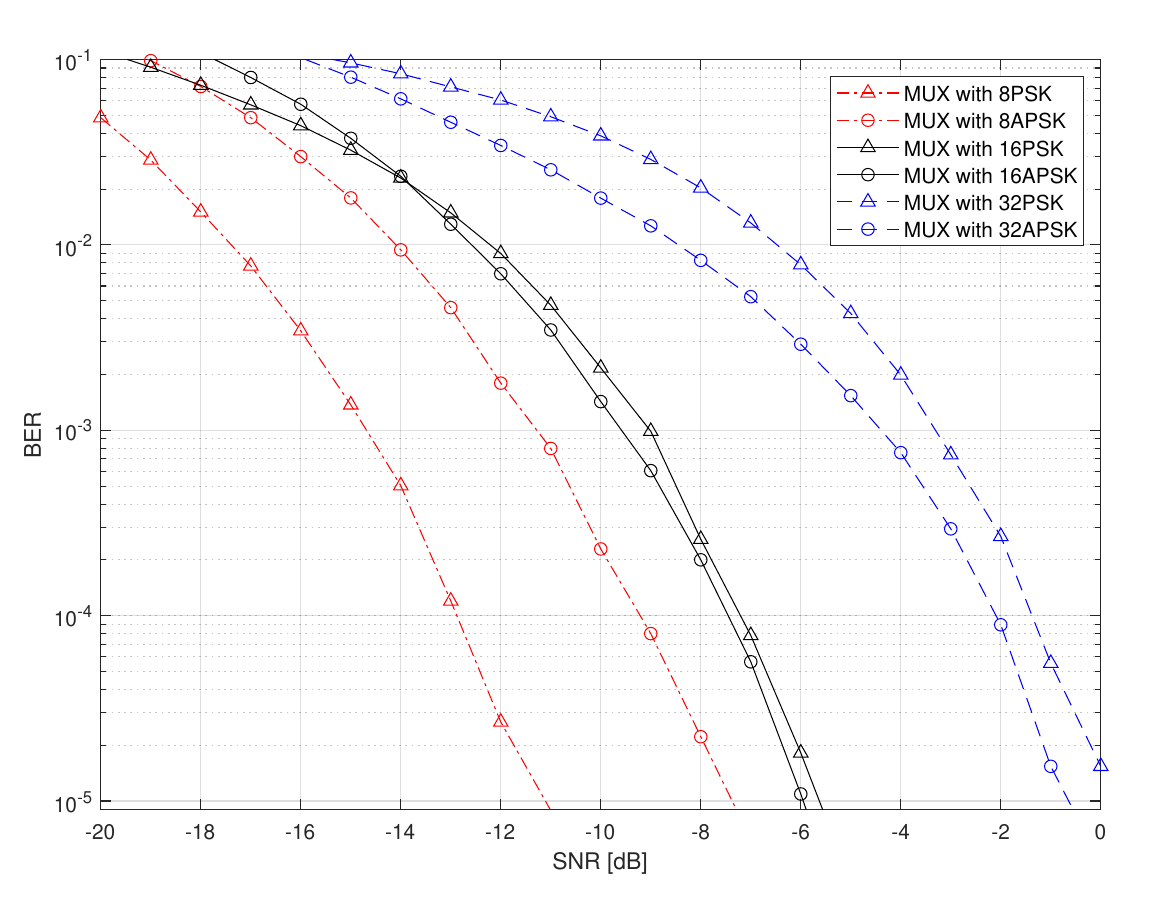}}
    \caption{The BER comparison results of the RIS-RGSM MUX with APSK and PSK for $N_R=5$, $N_a=2$ and $N=64$.}
    \label{fig:BER3}
\end{figure}

Finally, the performance of the RIS-RGSM MUX scheme using APSK and MUX scheme using PSK are compared, as shown in \figref{fig:BER3}. It can be found that for low modulation order of $M=8$, the performance of MUX with PSK scheme is better than that of MUX with APSK scheme. But when $M=16$, APSK scheme is slightly better than PSK scheme. The performance gain of APSK scheme is further expanded by increasing the modulation order to $M=32$. This is because when the modulation order is low, the equivalent received symbols of PSK scheme are sparsely distributed on the PSK ring and easy to distinguish. Meanwhile, in the APSK scheme, the number of activated elements is obtained from the mapping of information bits. For the equivalent received symbols on the inner ring of APSK, it is more difficult to distinguish them from other symbols or interference due to the power limitation. However, when the modulation order is high, the symbols on the single ring of PSK become dense and difficult to be distinguished. At this time, APSK can alleviate this problem as there is not only a single ring.

\section{Conclusion}  \label{v}
The RIS-RGSM scheme proposed in this paper is divided into diversity and MUX schemes. The RIS-RGSM diversity scheme is realized based on the state-of-the-art RIS-RGSSK scheme and a novel RIS-RGSM MUX scheme is proposed. In the MUX scheme, RIS controllers adjust the reflection phases of elements to simultaneously achieve phase modulation and the selection of receive antennas. Furthermore, the amplitudes of the received symbols are modulated by adjusting the on/off states of elements. The proposed MUX scheme can be easily extended to various modulation constellations. 
Then, the theoretical BER is derived which fits well with the simulation results. According to the simulation results, the proposed RIS-RGSM MUX scheme offers 3, 10, and 15 dB BER gains compared to the diversity scheme when $R=$7, 9, and 11 bpcu, respectively. Compared with the existing RIS-RGSSK scheme, the proposed RIS-RGSM scheme can significantly increase the transmission rate when the number of receive antennas is limited, and the RIS-RGSM scheme can maintain a good performance at the same transmission rate. In addition, for low modulation orders, the RIS-RGSM MUX with PSK scheme performs better than MUX with APSK scheme, while for high modulation orders, the APSK scheme is better than the PSK scheme. The proposed RIS-RGSM scheme enables more efficient communication.



\bibliographystyle{IEEEtran}
\bibliography{IEEEabrv,mybib}

\end{document}